\documentclass[%
 reprint,
 amsmath,amssymb,
 aps,
 nofootinbib,
]{revtex4-2}

\usepackage{graphicx}
\usepackage{dcolumn}
\usepackage{bm}
\usepackage{footnotehyper}

\begin{document}

\preprint{APS/123-QED}

\title{Analytical Solution of the Nonlinear Boltzmann Equation For Multicomponent Systems}

\author{Linyuan Wei}
\affiliation{International Joint Institute of Tianjin University, Fuzhou,  Tianjin University, Tianjin  300072, China}

\author{Yi Wang}
 \affiliation{Department of Physics, Tsinghua University, Beijing 100084, China}

\author{Jin Hu}
 \email{hu-j23@fzu.edu.cn}
\affiliation{Department of Physics, Fuzhou University, Fujian 350116, China}
 
\author{Baoyi Chen}
 \email{baoyi.chen@tju.edu.cn}
\affiliation{Department of Physics, Tianjin University, Tianjin 300354, China}


\date{\today}

\begin{abstract}

We present exact analytical solutions to the nonlinear relativistic Boltzmann equation for homogeneous, isotropic massless multi-component systems with non-isotropic scattering cross sections. 
For a two-component system, we identify three distinct classes of exact solutions, each corresponding to specific constraints among initial energy densities, particle number densities, and  scattering cross sections. These solutions comprise the equilibrium state and two novel classes: a hybrid structure featuring a non-equilibrium Maxwell–Jüttner distribution coupled with a nontrivial distribution; the former is unprecedented in  kinetic theory, while the latter refers to nontrivial dynamics shared by both species, which is novel in the relativistic context.
\end{abstract}

\maketitle

\section{Introduction}
The Boltzmann equation stand as a cornerstone of non-equilibrium statistical mechanics \cite{Hubert_2021}, connecting microscopic collision dynamics to macroscopic transport phenomena across diverse physical domains, such as quark–gluon plasma in relativistic heavy-ion collisions \cite{Xu:2004mz,PhysRevC.94.014909,Yao:2018zrg,Xing:2021xwc,Hebenstreit:2011pm,Wang:2021oqq,Palni:2024wdy}, neutron transport in nuclear reactors \cite{Bocanegra2020}, charge carrier transport in nanoscale transistors \cite{Martinez2020}, and neutrino transport in core-collapse supernovae \cite{Mezzacappa:1993gm}. While numerical methods, e.g.,  the Direct Simulation Monte Carlo (DSMC) and the Lattice Boltzmann Method (LBM), have achieved remarkable success \cite{Simeoni:2022hjh,10.1063/5.0268025}, analytical solutions remain invaluable for their profound theoretical significance, broad applicability and transparency in revealing how microscopic interactions drive entropy production, nonlinear relaxation, and the emergence of hydrodynamic behavior \cite{DeGroot:1980dk}.
Exact analytical solutions of the Boltzmann equation are exceedingly rare due to its nonlinear integro-differential structure. The celebrated BKW (Bobylev, Krook, and Wu)  solution \footnote{This solution was first derived  by R. S. Krupp in his master's thesis in 1967 \cite{dissertation}, predating its later rediscovery as reviewed in \cite{exactreview}.} for non-relativistic Maxwell molecules \cite{Bobylev1,KW,KW2} and recent breakthroughs for single-component relativistic systems \cite{Bazow:2015dha,Bazow:2016oky,Hu:2024utr,Wang:2025wyh} represent notable exceptions. However, physically realistic systems, such as quark–gluon plasma or early-universe particle mixtures, inherently consist of multiple interacting species, where inter-species scattering introduces additional complexity for  exact analytical treatment.

In this work, we bridge this gap by deriving exact analytical solutions for multi-component relativistic Boltzmann equations. Our framework systematically incorporates momentum-independent yet angle-dependent  cross sections, thereby capturing essential features of realistic microscopic interactions beyond the hard-sphere model employed in  previous works to solve for exact solutions, e.g., \cite{Bazow:2015dha,Bazow:2016oky}. For the specific two-component case, our main results comprise three classes of solutions: the first corresponds to the  equilibrium state; the second features both particle species following BKW-type distributions; and the third represents a hybrid structure combining a non-equilibrium Maxwell–J\"{u}ttner distribution with a BKW-type distribution. Notably, the second class of solutions is novel in relativistic kinetic theory, while the third class is novel in both relativistic and non-relativistic kinetic theory literature.

Below, we adopt the natural units \(k_B = c = \hbar =1\), the metric tensor \(g^{\mu \nu} = \text{diag}(1,-1,-1,-1)\), and the Lorentz-invariant integral measure for on-shell massless particles \(\int \mathrm{dP}_k \equiv \frac{2}{(2 \pi) ^3} \int d^4 p_k \theta(p_k ^0) \delta(p_k ^2)\).
~\\
\section{Multi-component Boltzmann Equation}\label{Boltzmann Equation}
For a mixture of $N$ particle species, the distribution function $f_k(x^\mu,p_k^\mu)$ of each component $k=1,2,\ldots,N$ obeys the relativistic Boltzmann equation
\begin{equation}
p_k ^{\mu} \partial_{\mu} f_k (x^{\mu}, p_k ^{\mu}) = \sum_{l=1} ^{N} C_{kl} (x^{\mu},p_k ^{\mu}), \label{eq:Boltzmann}
\end{equation}
with the collision term
\begin{align}
     C_{kl} (x^{\mu},p_k ^{\mu}) &\equiv \gamma_{kl} \int \mathrm{dP}_l  \mathrm{dP}_k ^\prime  \mathrm{dP}_l ^\prime (f_k (x^\mu , {p_k ^\prime } ^\mu) f_l (x^\mu, {p_l ^\prime } ^\mu) \notag \\
&- f_k(x^\mu , p_k ^\mu) f_l (x^\mu , p_l ^\mu)) W_{kl} (p_k ,p_l | p_k ^\prime , p_l ^\prime ),\label{eq:collition}
\end{align}
where $\gamma_{kl}=1-\frac{1}{2}\delta_{kl}$ ensures that $W_{kl}$ applies to both identical and distinct particles \cite{DeGroot:1980dk,Hu:2022vph}. The transition rate is $W_{kl}=(2\pi)^6s\sigma_{kl}(s,\Theta)\delta^4(p_k+p_l-p_k'-p_l')$, with $s$ the Mandelstam variable defined as   \(s \equiv (p_k ^\mu + p_l ^\mu ) (p_{k,\mu} + p_{l,\mu})\) and $\Theta$ the scattering angle in the center-of-momentum frame. We neglect external forces and quantum statistical effects, and assume $\sigma_{kl}=\sigma_{lk}$ by detailed balance. Besides,  we focus on elastic scattering processes whose scattering cross sections are independent of $s$.

To proceed, we further assume spatial homogeneity and isotropy. By introducing dimensionless variables $\tau\equiv T_0t$, $\hat{p}_k^\mu\equiv p_k^\mu/T_0$, and \(\hat{\sigma}_{kl} (\cos{\Theta}) \equiv T_0 ^2  \sigma_{kl} (\cos{\Theta})\) with $T_0$ a representative energy scale, we can cast Eq.~\eqref{eq:Boltzmann} into
\begin{align}
       \hat{p}_k ^{0} \partial_{\tau} f_k (\tau, \hat{p}_k^0) &= \sum_{l=1} ^{N} C_{kl} (\tau,\hat{p}_k^0) ,\label{eq:Bzm}  
\end{align}
with the collision term 
\begin{align}
      C_{kl} (\tau,\hat{p}_k^0) &= (2 \pi)^6  \gamma_{kl} \int  \mathrm{dP}_l \mathrm{dP}_k ^\prime \mathrm{dP}_l ^\prime  s  \hat{\sigma}_{kl}(\cos{\Theta}) \notag \\
      & \quad \times \delta ^4 (\hat{p_k} + \hat{p_l} - \hat{p_k} ^\prime - \hat{p_l} ^\prime) (f_k^\prime f_l^\prime - f_k f_l),\label{FC_kl}
\end{align}
where $f_k'\equiv f_k(\tau,p_k^{\prime0})$, and the hats on dimensionless quantities are omitted hereafter.
~\\

\section{Moment Method}\label{Moment Method}
We adopt the moment method to solve the nonlinear Boltzmann equation \cite{Hu:2024utr}. The scalar energy moment is defined as
\begin{equation}
\rho_k ^{(n)}(\tau) \equiv \int \mathrm{dP}_k (p_k ^0)^{n+1} f(\tau, p_k ^0),\label{scalar energy moment}
\end{equation}
where $n$ takes the integer value. The Boltzmann equation Eq.~\eqref{eq:Bzm} is transformed into 
\begin{equation}
    \partial_\tau  \rho_k ^{(n)}(\tau) = \sum_{l=1} ^N C_{kl} ^{(n)} (\tau) ,\label{JBEQ}
\end{equation}
with
\begin{align}
      C_{kl} ^{(n)}(\tau) &= (2 \pi)^6  \gamma_{kl} \int  \mathrm{dP}_k \mathrm{dP}_l \mathrm{dP}_k ^\prime  \mathrm{dP}_l ^\prime   s (p_k ^{0})^n \sigma_{kl}(\cos{\Theta})  \notag \\
    &\quad \times  \delta ^4 (p_k + p_l - p_k ^\prime  - p_l ^\prime ) (f_k^\prime f_l^\prime  - f_k f_l).\label{JC_kl}
\end{align}
The lowest two moments correspond to physical quantities: $\rho_k^{(0)}$ is the particle number density and $\rho_k^{(1)}$ is the energy density. Conservation laws impose
\begin{align}
\partial_\tau\rho_k^{(0)}=0,\quad
\partial_\tau \sum_{k=1}^N\rho_k^{(1)}=0,\label{eq:conserv}
\end{align}
where the vanishing of the right-hand side of the moment equations directly reflects the collision invariance of the relativistic Boltzmann equation. We impose the initial conditions $n_k=\rho_k^{(0)}(0)$, $e_k=\rho_k^{(1)}(0)$, where $n_k,e_k$ denote  the initial particle number density and energy density, respectively. The conservation laws under spatial homogeneity give $n_k=\rho_k^{(0)}(\tau)$ and $\sum_{k=1}^N e_k=\sum_{k=1}^N\rho_k^{(1)}(\tau)$.
~\\

\section{Exact  analytical solution}\label{analytical solution}
To be specific, we focus on the case \( N = 2 \). The  evolution of the system is governed by
\begin{equation}
    \partial_\tau  \rho_k ^{(n)}(\tau) = \sum_{l=1} ^2 C_{kl} ^{(n)} (\tau) , \quad k = 1, 2. \label{eq:transport}
\end{equation}

We adopt a class of  ansatz distribution functions (for various $m$) with the following parametrization:
\begin{align} \begin{split} f_k(\tau, p_k^0) = e^{- \frac{p_k^0}{\alpha(\tau)}} \sum_{i=0}^m A_{i,k}(\tau)  (p_k^0)^i, \quad k =1,2,\label{ansatz} \end{split} \end{align}
\noindent where $A_{i,k}(\tau)$  are time-dependent coefficients to be determined. This form is motivated by three considerations: (i) separability of $\tau$ and $p^0$
  dependence for analytical tractability; (ii) regularity for ensuring finite kinetic moments; and (iii) minimal complexity --- expand the $p^0$-dependent part as a power series in $p^0$ while employing an exponentially decaying regulating function. It can be readily verified that only two values of $m$ yield analytically tractable ansatz functions  \cite{Wang:2025wyh}: $m=0$ corresponds to the well-known Maxwell–J\"{u}ttner form, while $m=1$ gives rise to a nontrivial form. Focusing on the nontrivial case of $m = 1$, we are led to
\begin{align}
    f_k (\tau , p_k ^0) &=  e^{-\frac{p_k ^0}{\alpha_k (\tau)}} (A_k (\tau) + B_k (\tau) p_k ^0), \quad k =1,2,\label{fk}
\end{align}
where we have relabeled the coefficients as $A_{0,k}\rightarrow A_k$ and $A_{1,k}\rightarrow B_k$.

Substitute the ansatz distribution functions Eq.~\eqref{fk} into the left-hand sides of Eq.~\eqref{scalar energy moment}, and perform the integration to obtain
\begin{align}
    \rho_k ^{(n)} & = \frac{ \Gamma(n+3) \alpha_k (\tau)^{n+3} (A_k (\tau) + (n+3) \alpha_k (\tau) B_k (\tau))}{2 \pi ^2} . \label{rhok}
\end{align}
Similarly, integrating Eq.~\eqref{JC_kl} results in
\begin{align}
    C_{kl} ^{(n)} & = \frac{\gamma_{kl}}{2^n (2\pi)^6} \sum_{m=0} ^{n} C_n ^m  \big(\frac{A_l(\tau) B_k(\tau) - A_k(\tau) B_l(\tau)}{2}  \notag \\
    &\quad \times (J_{kl}^{(n-m,m,1)}- J_{kl}^{(n-m,m+1,0)}) \notag \\
    &\quad - \frac{B_k(\tau) B_l (\tau)}{4} (J_{kl}^{(n-m,m,2)}- J_{kl}^{(n-m,m+2,0)}) \big) ,\label{TC_kl}
\end{align}
with
\begin{align}
    J_{kl} ^{(b,d,e)} &= \frac{2^5 \pi^3 \alpha(\tau)^{b+d+e+6}}{(d+e+3)(d+e+5)}  \Gamma(b+d+e+6) \notag \\
    &\quad \times \sum_{g=0} ^{\min(d, e)}( \sigma_{kl} ^{(g)}  \frac{ \big(1 + (-1)^{d-g} \big) \big(1 + (-1)^{e-g} \big) }{4}  \notag \\
    &\quad \times \frac{d! e! }{(d-g)!!(d+g+1)!! (e-g)!! (e+g+1)!!} ), \label{J}
\end{align}
where $\alpha_k(\tau)=\alpha_l(\tau)=\alpha(\tau)$ is imposed as also noted in \cite{PhysRevLett.38.991}. Allowing $\alpha_k$ and $\alpha_l$ to evolve independently leads to analytically intractable coupled nonlinear equations, therefore, to ensure analytical solvability  and closure of the dynamical system, we impose the constraint  $\alpha_k(\tau)=\alpha_l(\tau)=\alpha(\tau)$. Here \(\sigma_{kl}^{(g)}\) represents the coefficients of the expansion of \(\sigma_{kl}(\cos{\Theta})\) in terms of Legendre polynomials
\begin{align}
    \sigma_{kl}(\cos{\Theta}) &= \sum_{g=0} ^\infty \sigma_{kl}^{(g)} P_g (\cos{\Theta}),\label{sigma}
\end{align}
where 
\begin{align}
    \sigma_{kl}^{(g)} \equiv \frac{2g+1}{2} \int_{-1} ^1 dx \sigma_{kl}(x) P_g (x),
\end{align}
and $P_g (x)$ denotes the Legendre polynomial. Further computational details can be  found in Supplemental Material and Ref.~\cite{DeGroot:1980dk}, pp. 375–380. It can be verified that Eq.~\eqref{TC_kl} manifestly preserves the collision invariance.

From Eqs.~\eqref{eq:conserv} and ~\eqref{rhok}, we get 
\begin{align}
    A_k(\tau ) &= \frac{n_k \pi^2}{\alpha(\tau)^3} - 3 B_k(\tau ) \alpha(\tau) ,\quad \label{A_k}
\end{align}
\begin{align}
    B_1(\tau) = -B_2(\tau) + \frac{\pi^2(e_1 + e_2 - 3 (n_1 + n_2) \alpha(\tau)) }{3 \alpha(\tau)^5}.\label{B_1}
\end{align}
The problem thus reduces to solving for $B_2(\tau)$ and $\alpha(\tau)$.

For \( n = 1 \), in addition to  the conservation law of the total energy density,  \( f_1 \) and \( f_2 \) must separately satisfy the transport equations within Eq.~\eqref{eq:transport}. First, we solve Eq.~\eqref{eq:transport} for \( k = 1 \) to obtain \( B_2 \), yielding
\begin{align}
    B_2 (\tau) &= \frac{\pi^2 }{3 (n_1 + n_2 ) \alpha(\tau) ^5} \big((e_1 +e_2) n_2  \notag \\
    &\quad + e^{-\frac{2}{3}(n_1 +n_2) \pi \tau (3\sigma_{12}^{(0)} - \sigma_{12}^{(1)})} (e_2 n_1 - e_1 n_2) \notag \\
    &\quad - 3 n_2 (n_1 + n_2) \alpha(\tau) \big).\label{TB_2}
\end{align}
Next, we determine $\alpha(\tau)$ from the second-order scalar moment equation ($n=2$). Solving Eq.~\eqref{eq:transport} for $k=1$ with $n=2$ yields
\begin{align}
    \alpha(\tau) &=  \frac{e_1 + e_2}{ 3  (n_1 + n_2)} \notag \\
    &\quad +\frac{ e^{-\frac{2}{3}(n_1 +n_2) \pi \tau (3\sigma_{12}^{(0)} - \sigma_{12}^{(1)})}(e_1 n_2- e_2 n_1)    }{ 3 n_1 (n_1 + n_2)} , \label{Talpha1}
\end{align}
or
\begin{align}
     &\alpha(\tau ) =  \frac{e_1 +e_2}{3(n_1 +n_2)}- \frac{e^{-\frac{2}{3} (n_1 +n_2 ) \pi \tau (3\sigma_{12}^{(0)} - \sigma_{12} ^{(1)})}}{3(n_1 +n_2)} \notag\\
     &\quad \times
     (e_2 n_1 - e_1 n_2)(5\sigma_{11} ^{(0)} - \sigma_{11}^{(2)} -  10\sigma_{12}^{(0)} + 2\sigma_{12}^{(2)})  \notag \\
     &\quad \times\big(n_1( 5\sigma_{11}^{(0)}- \sigma_{11}^{(2)} -30\sigma_{12}^{(0)}+10\sigma_{12}^{(1)})  \notag\\
    &\quad - 2 n_2 (10\sigma_{12}^{(0)} -5\sigma_{12}^{(1)} +\sigma_{12}^{(2)}) \big)^{-1} \notag \\
    &\quad + C e^{-\frac{1}{15} \pi \tau ( n_1 (5\sigma_{11}^{(0)} - \sigma_{11}^{(2)}) +2 n_2 (5\sigma_{12}^{(0)} - \sigma_{12}^{(2)}))} , \label{Talpha2}
\end{align}
where $C$ is an integration constant independent of $\tau$. However, this is not the end of the story: the $\alpha(\tau)$ obtained above must also self-consistently satisfy the transport equation for $k=2$ in Eq.~\eqref{eq:transport}, which yields additional solvability conditions. Specifically, Eq.~\eqref{Talpha1} solves Eq.~\eqref{eq:transport} for $k=2$ only when one of the following conditions is satisfied
\begin{align}
    n_2 &= \frac{e_2 n_1}{e_1} ,\label{Condition 1}\\
    \sigma_{12}^{(1)} &=  \frac{1}{10} (30\sigma_{12}^{(0)} - 5 \sigma_{22}^{(0)} + \sigma_{22}^{(2)}).\label{Condition 2}
\end{align}
Similarly, consistency of Eq.~\eqref{Talpha2} with the $k=2$ transport equation in Eq.~\eqref{eq:transport}, together with the requirement that $C$ is independent of $\tau$,  demands that at least one of the following conditions be satisfied
\begin{align}
      n_2 &= \frac{e_2 n_1}{e_1} ,\label{Condition 3}\\
    \sigma_{12}^{(1)} &=  \frac{1}{10} (30\sigma_{12}^{(0)} - 5 \sigma_{11}^{(0)} + \sigma_{11}^{(2)}),\label{Condition 4}\\
     \sigma_{12}^{(1)} &= -\big(-500{\sigma_{12}^{(0) }}^2 + 80 \sigma_{12}^{(0)} \sigma_{12}^{(2)} +4 {\sigma_{12}^{(2)}}^2 + 150 \sigma_{12}^{(0)} \sigma_{22}^{(0)} \notag \\
     &\quad-30 \sigma_{12}^{(0)} \sigma_{22}^{(2)} + (5 \sigma_{11}^{(0)}-\sigma_{11}^{(2)}) (30 \sigma_{12}^{(0)} -5 \sigma_{22}^{(0)} \notag \\
     &\quad+\sigma_{22}^{(2)} ) \big)/ \big(10(-5 \sigma_{11}^{(0)}  + \sigma_{11}^{(2)} +20 \sigma_{12}^{(0)} -4  \sigma_{12}^{(2)}\notag \\
     &\quad -5 \sigma_{22}^{(0)} +  \sigma_{22}^{(2)}) \big). \label{Condition 5}
\end{align}
Through this procedure, all free parameters are uniquely determined, yielding complete solutions to the multi-component Boltzmann equations. It is noteworthy that the derivation strategically utilized only the moment equations for $n=1$ and $n=2$. However, the solutions must, in principle, satisfy moment equations of all orders. Consistency checks confirm that the results obtained here indeed satisfy the moment equations of arbitrary order.


 Thus, we obtain the following four sets of exact analytical solutions to the multi-component nonlinear relativistic Boltzmann equation, together with the precise conditions under which they are physically sensible:
  \begin{itemize}
     \item Case 1 - Eq.~\eqref{fk} with  \( A_k(\tau) \), \( B_1(\tau) \), \( B_2(\tau) \) given by Eqs.~\eqref{A_k}, \eqref{B_1}, \eqref{TB_2} and \( \alpha(\tau) \) specified by Eq.~\eqref{Talpha1} (or Eq.~\eqref{Talpha2}), together with Condition \eqref{Condition 1},
     \item Case 2 -Eq.~\eqref{fk} with  \( A_k(\tau) \), \( B_1(\tau) \), \( B_2(\tau) \) given by Eqs.~\eqref{A_k}, \eqref{B_1}, \eqref{TB_2} and \( \alpha(\tau) \) specified by Eq.~\eqref{Talpha1}, together with Condition \eqref{Condition 2},
     \item Case 3 - Eq.~\eqref{fk} with  \( A_k(\tau) \), \( B_1(\tau) \), \( B_2(\tau) \) given by Eqs.~\eqref{A_k}, \eqref{B_1}, \eqref{TB_2} and \( \alpha(\tau) \) specified by Eq.~\eqref{Talpha2}, together with Condition \eqref{Condition 4} , 
     \item Case 4 -  Eq.~\eqref{fk} with  \( A_k(\tau) \), \( B_1(\tau) \), \( B_2(\tau) \) given by Eqs.~\eqref{A_k}, \eqref{B_1}, \eqref{TB_2} and \( \alpha(\tau) \) specified by Eq.~\eqref{Talpha2}, together with Condition \eqref{Condition 5}.
 \end{itemize}
 Note that Case 2 and Case 3 are related by $1\leftrightarrow 2$ symmetry and we shall focus on Case 2 hereafter. It can be shown that $B_1=0$ for Case 2, $B_2=0$ for Case 3 and $B_1=B_2=0$ for Case 1. In the following  section, we will present a detailed comparative analysis and discussion of the solutions for the four cases.
 
As a final consistency check, the multi-component solutions reduce to their single-component counterparts under the substitutions $\sigma_{ij}^{(g)} \to \sigma^{(g)}$, $n_1+n_2 \to n_0$, $e_1+e_2 \to e_0$, and $\gamma_{kl} \to 1/2$. Specifically, under Condition~\eqref{Condition 4} or \eqref{Condition 5}, $\alpha(\tau)$ becomes
\begin{align}
    \alpha (\tau) = \frac{e_0}{3 n_0} + \left(\frac{e_2}{3n_2} - \frac{e_0}{3 n_0}\right) 
    e^{-\frac{1}{15} n_0 \pi \tau (5 \sigma^{(0)} - \sigma^{(2)})}, \label{single_alpha 1}
\end{align}
or
\begin{align}
    \alpha(\tau) =\frac{e_0}{3 n_0} + (\frac{e_1 - e_2}{3 (n_1 -n_2)} - \frac{e_0}{3 n_0}  ) e^{-\frac{1}{15} n_0 \pi \tau (5 \sigma^{(0)} - \sigma^{(2)})},\label{single alpha 3}
\end{align}
which coincides with  Eq.~(3.12) of Ref.~\cite{Hu:2024utr}, with $\frac{e_2}{3n_2}$ or $\frac{e_1 - e_2}{3(n_1 -n_2)}$  acting as the free initial data  therein. A similar correspondence holds for $\alpha(\tau)$ under Condition~\eqref{Condition 2}.  

~\\

\section{Discussion}\label{Discussion}
In this section, we present a comparative analysis and physical discussion of the four classes of solutions derived above (Case 1 to 4). We examine the relationships among these solutions and their asymptotic behavior. Furthermore, we delineate the physically admissible parameter regimes, i.e., the ranges of parameters for which these solutions qualify as physically meaningful.

Regarding Case 1, both species exhibit equilibrium distribution functions, which represents a trivial solution since the equilibrium state is by construction an exact solution of the Boltzmann equation and the inevitable fate of its time evolution. Turning to Cases 2 and 3, they are related by $1\leftrightarrow 2$ exchanging symmetry; thus we consider the former one. From Eqs.~\eqref{Condition 2} and \eqref{Talpha1}, it can be seen that the evolution of $f_1$ and $f_2$ is governed by $\sigma_{22}$ and $\sigma_{12}$.  At first glance, this appears puzzling, as the evolution of species 1 is independent of $\sigma_{11}$, i.e., its self-scattering cross-section. However, a more careful analysis reveals that although $f_1$
  exhibits explicit time dependence, it retains the standard Maxwell–J\"{u}ttner form at all times ($B_1=0$), and consequently the collision integral $C_{11}$
  vanishes identically, indicating that self-scattering no longer affects the evolution of $f_1$
  at this stage. Nevertheless, species 1 continues to interact with species 2, which accounts for the residual time dependence in its distribution; in contrast, the distribution of species 2 takes the standard BKW form. Therefore, Cases 2 and 3 feature a hybrid structure in which one component follows a non-equilibrium Maxwell–J\"{u}ttner distribution while the other follows a BKW-type distribution — a result that, to the best of our knowledge, represents a novel finding for both relativistic and non-relativistic kinetic theory. Finally in Case 4, both species exhibit BKW-type distributions. One may wonder whether Cases 2 and 3 represent intermediate stages in the evolution of Case 4 — specifically, whether $B_1$ or $B_2$ vanishes prematurely during the dynamical evolution. However, a more careful analysis reveals that $B_k(\tau \to \infty)=0$ for $k=1,2$, with the relaxation governed by the same exponential function in Case 4. 
  This implies that Cases 2 and 3 do not represent intermediate stages of Case 4 evolution, but rather constitute a distinct class of solutions.

  Besides, our results provide an analytical example demonstrating the decisive role of initial conditions and scattering cross-sections in determining the system's dynamical trajectory. For the flexible ansatz distribution function in Eq.~\eqref{ansatz} (or equivalently Eq.~\eqref{fk}), equilibrium solutions arise when the initial data satisfy Condition~\eqref{Condition 1}. Nontrivial time evolution emerges only when Condition~\eqref{Condition 1} is violated; in this case, the system follows distinct dynamical trajectories depending on whether the scattering cross-sections satisfy Condition~\eqref{Condition 2}, \eqref{Condition 4}, or \eqref{Condition 5}, which are analytically described by Cases 2, 3, and 4, respectively.

  Next, we turn to the asymptotic behavior of these analytical solutions.  The distribution functions in Cases 2, 3, and 4 all asymptotically approach the same equilibrium distribution as the system evolves over sufficiently long times, i.e.,
\begin{align}
    \lim_{\tau \rightarrow  \infty} f_1 (\tau ,p_1 ^0)&=  \frac{27 e^{-\frac{3(n_1 +n_2) p_1 ^0 }{e_1 +e_2} } n_1 (n_1 +n_2)^3 \pi^2}{(e_1 +e_2)^3 }, \label{equilibrium solution 3}\\
      \lim_{\tau \rightarrow  \infty} f_2 (\tau ,p_2 ^0)&=  \frac{27 e^{-\frac{3(n_1 +n_2) p_2 ^0 }{e_1 +e_2} } n_2 (n_1 +n_2)^3 \pi^2}{(e_1 +e_2)^3 } \label{equilibrium solution 4}.
\end{align}
It can be readily shown that, in the final equilibrium state, the two components reach thermodynamic equilibrium characterized by a common temperature $\frac{ e_1+ e_2}{3(n_1 + n_2)}$. We emphasize that Eqs.~\eqref{equilibrium solution 3} and \eqref{equilibrium solution 4} are consistent with Case 1  under the condition
 $ n_2 = \frac{e_2 n_1}{e_1}$.

 Finally, we discuss the physically admissible parameter regimes for the initial data. The physicality conditions we impose are non-negativity and finiteness of the distribution functions. For Case 2, the aforementioned physical requirements lead to the following conditions
 \begin{align}
    \frac{e_2}{4 n_2} &\leq \frac{e_1}{3n_1} \leq \frac{e_2}{3 n_2}, \quad \sigma_{22}^{(2)} < 5 \sigma_{22}^{(0)}. \label{parameter sigma_22}
\end{align}
In the single-component limit, under the substitutions $\sigma_{ij}^{(g)} \to \sigma^{(g)}$, $n_1+n_2 \to n_0$, $e_1+e_2 \to e_0$, setting \(\sigma^{(0)} = 1/(2\pi)\) reduces Eq.~\eqref{parameter sigma_22} to
 \begin{align}
    \frac{e_0}{4 n_0} &\leq \frac{e_1}{3n_1} \leq \frac{e_0}{3 n_0}, \quad \sigma^{(2)} < \frac{5}{2\pi}, \label{parameter isotropic}
\end{align}
which consistently recovers Eqs.~(4.1) and (4.2) in Ref.~\cite{Hu:2024utr}.
Similar considerations apply to Case 4, though the resulting physical constraints are algebraically cumbersome and are not displayed here. Notably, the analytical solution reported in Refs.~\cite{Bazow:2015dha,Bazow:2016oky} corresponds to the special case of Eq.~\eqref{parameter isotropic} when the lower bound is saturated.
  
~\\

\section{Summary and outlook}\label{Summary}
In summary, we have derived exact analytical solutions to the nonlinear relativistic Boltzmann equation for homogeneous, isotropic, multi-component massless systems with non-isotropic scattering cross sections. By employing the moment method with a tailored ansatz distribution function, we identified distinct solution classes for a two-component system governed by precise constraints among initial energy densities, particle number densities, and scattering cross sections. These solutions include the  equilibrium state, novel hybrid structures combining Maxwell–Jüttner and BKW-type distributions, as well as pure BKW-type dynamics for both species. 

Future extensions could incorporate spatial inhomogeneities, momentum-space anisotropies, inelastic scattering processes, and nontrivial background fields, which are essential for realistic descriptions of quark–gluon plasma or other relativistic many-body systems. The analytical solutions derived here provide rigorous benchmarks for numerical algorithms and establish a theoretical foundation for systematic hydrodynamic modeling in multi-component relativistic systems. We expect that these results will deepen our understanding of relativistic kinetic theory and find applications across diverse domains, from early-universe cosmology to relativistic heavy-ion collisions.



\vspace{0.5cm}
\noindent {\bf Acknowledgments}:  This work was financially supported by the National Natural Science Foundation of China under Grant No.12505149. 

\bibliography{ref} 

@misc{Wang:2025wyh,
    author = "Wang, Yi and Zhao, Xuan and Xu, Zhe and Hu, Jin",
    title = "{Analytical Solution and Lie Algebra of the Relativistic Boltzmann Equation}",
    eprint = "2511.08652",
    archivePrefix = "arXiv",
    primaryClass = "nucl-th",
    month = "11",
    year = "2025"
}

@article{Bazow:2015dha,
    author = "Bazow, D. and Denicol, G. S. and Heinz, U. and Martinez, M. and Noronha, J.",
    title = "{Analytic solution of the Boltzmann equation in an expanding system}",
    eprint = "1507.07834",
    archivePrefix = "arXiv",
    primaryClass = "hep-ph",
    reportNumber = "INT-PUB-15-038",
    doi = "10.1103/PhysRevLett.116.022301",
    journal = "Phys. Rev. Lett.",
    volume = "116",
    number = "2",
    pages = "022301",
    year = "2016"
}

@article{Xu:2004mz,
	author = "Xu, Zhe and Greiner, Carsten",
	title = "{Thermalization of gluons in ultrarelativistic heavy ion collisions by including three-body interactions in a parton cascade}",
	eprint = "hep-ph/0406278",
	archivePrefix = "arXiv",
	doi = "10.1103/PhysRevC.71.064901",
	journal = "Phys. Rev. C",
	volume = "71",
	pages = "064901",
	year = "2005"
}

@book{DeGroot:1980dk,
	author = "De Groot, S. R.",
	editor = "Van Leeuwen, W. A. and Van Weert, C. G.",
	title = "{Relativistic Kinetic Theory. Principles and Applications}",
	year = "1980"
}

@article{Hu:2022vph,
	author = "Hu, Jin and Shi, Shuzhe",
	title = "{Multicomponent second-order dissipative relativistic hydrodynamics with binary reactive collisions}",
	eprint = "2204.10100",
	archivePrefix = "arXiv",
	primaryClass = "hep-ph",
	doi = "10.1103/PhysRevD.106.014007",
	journal = "Phys. Rev. D",
	volume = "106",
	number = "1",
	pages = "014007",
	year = "2022"
}

@article{PhysRevLett.38.991,
  title = {Exact Solution of Boltzmann Equations for Multicomponent Systems},
  author = {Krook, Max and Wu, Tai Tsun},
  journal = {Phys. Rev. Lett.},
  volume = {38},
  issue = {18},
  pages = {991--993},
  numpages = {0},
  year = {1977},
  month = {May},
  publisher = {American Physical Society},
  doi = {10.1103/PhysRevLett.38.991},
  url = {https://link.aps.org/doi/10.1103/PhysRevLett.38.991}
}

@phdthesis{Hebenstreit:2011pm,
    author = "Hebenstreit, Florian",
    title = "{Schwinger effect in inhomogeneous electric fields}",
    eprint = "1106.5965",
    archivePrefix = "arXiv",
    primaryClass = "hep-ph",
    school = "Graz U.",
    year = "2011"
}

@article{10.1063/5.0268025,
    author = {Silva, Goncalo and Ginzburg, Irina},
    title = {Lattice Boltzmann method simulation of rotating channel flows: Improved modeling of diffusion and advection terms},
    journal = {Physics of Fluids},
    volume = {37},
    number = {5},
    pages = {052010},
    year = {2025},
    month = {05},
    issn = {1070-6631},
    doi = {10.1063/5.0268025},
    url = {https://doi.org/10.1063/5.0268025},
  
}

@article{Simeoni:2022hjh,
    author = "Simeoni, Daniele and Gabbana, Alessandro and Succi, Sauro",
    title = "{Bjorken Flow Revisited: Analytic and Numerical Solutions in Flat Space-Time Coordinates}",
    eprint = "2202.09133",
    archivePrefix = "arXiv",
    primaryClass = "nucl-th",
    doi = "10.4208/cicp.OA-2022-0051",
    journal = "Commun. Comput. Phys.",
    volume = "33",
    number = "1",
    pages = "174--188",
    year = "2023"
}

@article{Wang:2021oqq,
    author = "Wang, Zeyan and Zhao, Jiaxing and Greiner, Carsten and Xu, Zhe and Zhuang, Pengfei",
    title = "{Incomplete electromagnetic response of hot QCD matter}",
    eprint = "2110.14302",
    archivePrefix = "arXiv",
    primaryClass = "hep-ph",
    doi = "10.1103/PhysRevC.105.L041901",
    journal = "Phys. Rev. C",
    volume = "105",
    number = "4",
    pages = "L041901",
    year = "2022"
}

@article{Mezzacappa:1993gm,
    author = "Mezzacappa, A. and Bruenn, S. W.",
    title = "{Type II supernovae and Boltzmann neutrino transport: The Infall phase}",
    doi = "10.1086/172394",
    journal = "Astrophys. J.",
    volume = "405",
    pages = "637--668",
    year = "1993"
}

@article{Martinez2020,
  author    = {Martinez, Antonio and Barker, John R.},
  title     = {Quantum Transport in a Silicon Nanowire FET Transistor: Hot Electrons and Local Power Dissipation},
  journal   = {Materials},
  volume    = {13},
  number    = {15},
  pages     = {3326},
  year      = {2020},
  doi       = {10.3390/ma13153326},
  url       = {https://doi.org/10.3390/ma13153326}
}

@article{Bocanegra2020,
  title={Lattice Boltzmann Method Applied to Nuclear Reactors—A Systematic Literature Review},
  author={Johan Augusto Bocanegra Cifuentes and Davide Borelli and Antonio Cammi and Guglielmo Lomonaco and Mario Misale},
  journal={Sustainability},
  year={2020},
  volume={12},
  number={18},
  pages={7835},
  doi={10.3390/su12187835},
  url={https://doi.org/10.3390/su12187835}
}

@article{Palni:2024wdy,
    author = "Palni, Prabhakar and others",
    title = "{Dynamics of hot QCD matter 2024~{\textemdash} Bulk properties}",
    eprint = "2412.10779",
    archivePrefix = "arXiv",
    primaryClass = "nucl-th",
    doi = "10.1142/S0218301325440021",
    journal = "Int. J. Mod. Phys. E",
    volume = "34",
    number = "07",
    pages = "2544002",
    year = "2025"
}

@article{Xing:2021xwc,
    author = "Xing, Wen-Jing and Qin, Guang-You and Cao, Shanshan",
    title = "{Perturbative and non-perturbative interactions between heavy quarks and quark-gluon plasma within a unified approach}",
    eprint = "2112.15062",
    archivePrefix = "arXiv",
    primaryClass = "hep-ph",
    doi = "10.1016/j.physletb.2023.137733",
    journal = "Phys. Lett. B",
    volume = "838",
    pages = "137733",
    year = "2023"
}

@article{Yao:2018zrg,
    author = {Yao, Xiaojun and Ke, Weiyao and Xu, Yingru and Bass, Steffen and M{\"u}ller, Berndt},
    editor = "Antinori, Federico and Dainese, Andrea and Giubellino, Paolo and Greco, Vincenzo and Lombardo, Maria Paola and Scomparin, Enrico",
    title = "{Quarkonium production in heavy ion collisions: coupled Boltzmann transport equations}",
    eprint = "1807.06199",
    archivePrefix = "arXiv",
    primaryClass = "nucl-th",
    doi = "10.1016/j.nuclphysa.2018.10.005",
    journal = "Nucl. Phys. A",
    volume = "982",
    pages = "755--758",
    year = "2019"
}

@article{PhysRevC.94.014909,
  title = {Linearized Boltzmann transport model for jet propagation in the quark-gluon plasma: Heavy quark evolution},
  author = {Cao, Shanshan and Luo, Tan and Qin, Guang-You and Wang, Xin-Nian},
  journal = {Phys. Rev. C},
  volume = {94},
  issue = {1},
  pages = {014909},
  numpages = {13},
  year = {2016},
  month = {Jul},
  publisher = {American Physical Society},
  doi = {10.1103/PhysRevC.94.014909},
  url = {https://link.aps.org/doi/10.1103/PhysRevC.94.014909}
}

@article{Hubert_2021,
   title={Understanding physics: ‘What?’, ‘Why?’, and ‘How?’},
   volume={11},
   ISSN={1879-4920},
   url={http://dx.doi.org/10.1007/s13194-021-00399-w},
   DOI={10.1007/s13194-021-00399-w},
   number={3},
   journal={European Journal for Philosophy of Science},
   publisher={Springer Science and Business Media LLC},
   author={Hubert, Mario},
   year={2021},
   month=aug }

@article{Hu:2024utr,
    author = "Hu, Jin",
    title = "{Analytical solution of the nonlinear relativistic Boltzmann equation}",
    eprint = "2411.16448",
    archivePrefix = "arXiv",
    primaryClass = "hep-ph",
    doi = "10.1007/JHEP07(2025)066",
    journal = "JHEP",
    volume = "2025",
    number = "07",
    pages = "066",
    year = "2025"
}

@article{Bobylev1,
	author = "Bobylev, A. V.",
	title = "{The Fourier transform method for the Boltzmann equation for Maxwell molecules}",
	journal = "Sov. Phys. Dokl",
	volume = "20",
	pages = "820-822",
	year = "1976",
}

@article{exactreview,
	author = "Ernst, Matthieu H.",
	title = "{Exact Solutions of the Nonlinear Boltzmann Equation}",
	journal = "J.Stat.Phys",
	volume = "34",
	pages = "1001-1017",
	year = "1984",
}

@article{dissertation,
	author = "Krupp, R.S.",
	title = "{A Nonequilibrium Solution of the Fourier Transformed Boltzmann Equation}",
	journal = "M.Sc. thesis, MIT",
	year = "1967",
}

@article{KW2,
	title = {Formation of Maxwellian Tails},
	author = {Krook, Max and Wu, Tai Tsun},
	journal = {Phys. Rev. Lett.},
	volume = {36},
	issue = {19},
	pages = {1107--1109},
	numpages = {0},
	year = {1976},
	month = {May},
	publisher = {American Physical Society},
	doi = {10.1103/PhysRevLett.36.1107},
}

@article{Bazow:2016oky,
	author = "Bazow, D. and Denicol, G. S. and Heinz, U. and Martinez, M. and Noronha, J.",
	title = "{Nonlinear dynamics from the relativistic Boltzmann equation in the Friedmann-Lema\^\i{}tre-Robertson-Walker spacetime}",
	eprint = "1607.05245",
	archivePrefix = "arXiv",
	primaryClass = "hep-ph",
	doi = "10.1103/PhysRevD.94.125006",
	journal = "Phys. Rev. D",
	volume = "94",
	number = "12",
	pages = "125006",
	year = "2016"
}

@article{KW,
	author = {Krook, Max and Wu, Tai Tsun},
	title = "{Exact solutions of the Boltzmann equation}",
	journal = {Phys. Fluids},
	volume = {20},
	number = {10},
	pages = {1589-1595},
	year = {1977},
	month = {10},
	doi = {10.1063/1.861780},
}
\bibliographystyle{apsrev4-2} 

\end{document}